\documentclass[aps,prl,twocolumn,superscriptaddress,showkeys,amsmath,amssymb,floatfix,secnumarabic]{revtex4}
\RequirePackage{lineno} 
\usepackage{booktabs,graphicx,mathrsfs,verbatim}
\usepackage[colorlinks=true,citecolor=blue,filecolor=blue,linkcolor=blue,urlcolor=blue]{hyperref}
\usepackage[usenames,dvipsnames]{color}
\usepackage{ulem}
\usepackage{color}

\newcommand{\columbia}{\affiliation{Physics Department, Columbia University, New York, NY 10027, USA}}
\newcommand{\rice}{\affiliation{Department of Physics and Astronomy, Rice University, Houston, TX 77005, USA}}
\newcommand{\munster}{\affiliation{Institut f\"ur Kernphysik, Westf\"alische Wilhelms-Universit\"at M\"unster, 48149 M\"unster, Germany}}
\newcommand{\purdue}{\affiliation{Department of Physics, Purdue University, West Lafayette, IN 47907, USA}}

\newcommand{\bchoiemail}{\email{bc2196@columbia.edu}}
\begin{document}

\title{Measurement of the Quantum Efficiency of Hamamatsu R8520 Photomultipliers at Liquid Xenon Temperature}

\author{E.~Aprile}\columbia
\author{M.~Beck}\munster
\author{K.~Bokeloh}\munster                                                                                                                                              
\author{R.~Budnik}\columbia                                                                                                                                        
\author{B.~Choi}\columbia\bchoiemail                                                                                                                                                   
\author{H.~A.~Contreras}\columbia                                                                                                                                                   
\author{K.-L.~Giboni}\columbia                                                                                                                                                      
\author{L.~W.~Goetzke}\columbia                                                                                                                                                                                                                                                                                                           
\author{R.~F.~Lang}\columbia\purdue                                         
\author{K.~E.~Lim}\columbia                                                                                                                       
\author{A.~J.~Melgarejo Fernandez}\columbia                                                                                                                                                                                                                                                                                           
\author{G.~Plante}\columbia                                                                                                                                                      
\author{A.~Rizzo}\columbia
\author{P.~Shagin}\rice 
\author{C.~Weinheimer}\munster


\begin{abstract}
Vacuum ultraviolet light sensitive photomultiplier tubes directly coupled to liquid xenon are being used to efficiently detect the 178 nm scintillation light in a variety of liquid xenon based particle detectors. Good knowledge of the performance of these photomultipliers under cryogenic conditions is needed to
properly characterize these detectors. Here, we report on measurements of the quantum efficiency of 
Hamamatsu R8520 photomultipliers, used in the XENON Dark Matter Experiments. The quantum efficiency measurements at room temperature
agree with the values provided by Hamamatsu. At low temperatures, between 160K and 170K, the
quantum efficiency increases by $\sim5-11$\% relative to the room temperature values.
\end{abstract}


\maketitle


\section{1. Introduction}
\label{}

Liquid Xenon (LXe) detectors are currently a favored choice for rare event search experiments, from dark matter direct detection to neutrinoless double beta decay~\cite{Aprile:2009dv}~\cite{Danilov:2000}~\cite{Ackerman:2011}~\cite{Abe:2009}. In particular, the XENON100 dark matter experiment~\cite{Aprile:2011}\cite{Aprile:2011hi} uses 242 Hamamatsu R8520-06-Al photomultiplier tubes (PMTs)~\cite{hamamatsu} to detect the scintillation light produced by Weakly Interacting Massive Particles (WIMPs) as they scatter off Xe nuclei. 

These compact metal-channel PMTs were specifically designed to work at 177 K, close to the temperature of LXe, and at a pressure up to 5 atm. They have a quartz window and a bialkali photocathode, for high sensitivity and low dark current in the UV regime \citep{Hamamatsu_handbook}. Recently, a new version of the same PMT (R8520-406) with improved sensitivity at 178 nm, the scintillation wavelength of LXe, was produced by Hamamatsu and procured for an upgrade of the XENON100 experiment. 

The quantum
efficiency (QE) is one of the most important characteristics of a PMT. The QE is defined as
the ratio between the number of photoelectrons emitted from the
photocathode and the number of incident photons. The energy absorbed from the incident photons is transferred to the valence band of the photocathode. Since the photoemission
process has a probabilistic nature, not all of the electrons that 
absorb energy from the incident photons are emitted as photoelectrons~\cite{Hamamatsu_handbook}.

A quantity related to the QE is the radiant sensitivity ($\mathrm{Sk}$), used to
express the spectral characteristics and especially the relationship
between the photocathode response and the incident light
wavelength. The radiant sensitivity is defined as the
photoelectric current generated by the photocathode, I$_\mathrm{K}$, divided by the
incident radiant flux at a given
wavelength, L$_\mathrm{P}$:

\begin{equation}
\mathrm{Sk} = \frac{\mathrm{I_K}}{\mathrm{L_P}}\\
\label{rad_sensitivity}
\end{equation}
The relationship between the QE and the
$\mathrm{Sk}$ is given by~\cite{Hamamatsu_handbook}

\begin{equation}
\mathrm{QE} = \frac{h\cdot c}{\lambda \cdot e}\cdot \mathrm{Sk} \simeq \frac{1240}{\lambda}\cdot \mathrm{Sk}\cdot 100 ~[\%]\\
\label{qe_sk}
\end{equation}
where $h$ is Planck's constant, $c$ is the speed of light in vacuum, $\lambda$ is the wavelength of the
incident light in nm, $e$ is
the electron charge, and $\mathrm{Sk}$ is the radiant sensitivity in A/W.

The radiant sensitivity is often provided by the
manufacturer for each PMT. More generally, the blue sensitivity index (sometimes called cathode blue sensitivity), $\mathrm{Sk_{b}}$, which
is the photoelectric current generated from the photocathode with a
blue filter interposed in the same setup, is specified. However, this value is usually measured only at room temperature. Hence, a
quantitative study of the QE as a function of temperature is
necessary for an improved understanding of LXe experiments.
 
\section{2. Experimental Setup}

\begin{figure}[htb]
\begin{center}
\includegraphics[width=9cm]{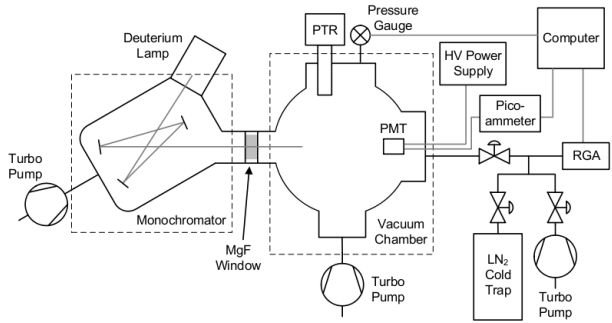}
\caption{A schematic view of the setup. LN$_{2}$ cold trap (cryopump) and the turbo pump system were installed to reduce the water level at different temperatures and pressures inside the vacuum chamber.}
\label{schematic}
\end{center}
\end{figure}

A schematic of the setup is shown in Figure~\ref{schematic}. It
consists of a deuterium lamp and two independent vacuum chambers. The first chamber contains a monochromator used to select a narrow wavelength range from
the spectrum of the deuterium lamp. The second chamber hosts the PMT being
tested.

A deuterium lamp (McPherson, Model 632) emits light between 115 nm and 400 nm and is particularly useful for measurements in the vacuum ultraviolet (VUV) and deep UV. The output spectrum between 380~nm and 165~nm is continuous~\citep{McPherson}. Below 165~nm, molecular lines predominate. The light emitted from
the lamp is transmitted by a MgF$_{2}$ window and passes through an entrance slit
into a vacuum monochromator (McPherson, Model 218) where a narrow wavelength range
from the lamp is selected. The monochromator is composed of a snap-in
diffraction grating that is optimized for wavelengths between 150~nm and 300~nm, and
two Al+MgF$_{2}$ coated mirrors that have a reflectivity of
$\sim$75\% at 178~nm~\citep{McPherson}. The diffraction grating can be remotely adjusted with a stepper motor control unit.  After the monochromator, the light with the
selected wavelength propagates through an exit
slit to the next vacuum chamber. The size of both the entrance and the exit slits can be adjusted
from 10~$\mu$m to 2~mm in order to modify the flux of the beam and the reciprocal linear dispersion. For this work, the size of both slits was set to 2~mm. The MgF$_{2}$
window that separates the
monochromator and the vacuum chamber has an optical transmissivity of $\sim$80\%~\citep{MgF2_transmission} for light with a wavelength of 178 nm. Independent vacuum pumping stations are
used to prevent impurities from the monochromator side, such as oil from the stepper motor, from reducing
the optical transparency of the PMT window. A Pfeiffer HiCube 80 Eco
pumping station on the monochromator and a Pfeiffer TMU071P 
turbomolecular pump with a dry backing pump on the second vacuum chamber keep the pressure at about 1.5$\times$10$^{-5}$~mbar and below 10$^{-6}$~mbar, respectively.
 
The vacuum chamber is equipped with a motion feedthrough that allows for linear motion in three dimensions as well as rotation. The photodetector --- either a PMT or a
photodiode for calibration purposes --- is mounted on this rotator, which can in addition be translated in both the horizontal and vertical directions. A customized collimator is installed on the port of the vacuum chamber where the light goes through. This reduces the size of the beam to $\sim\textless$1 cm and allows for more controlled and parallel beam shining on the photodetector.

\begin{figure}[htb]
\begin{center}
\includegraphics[width=9cm]{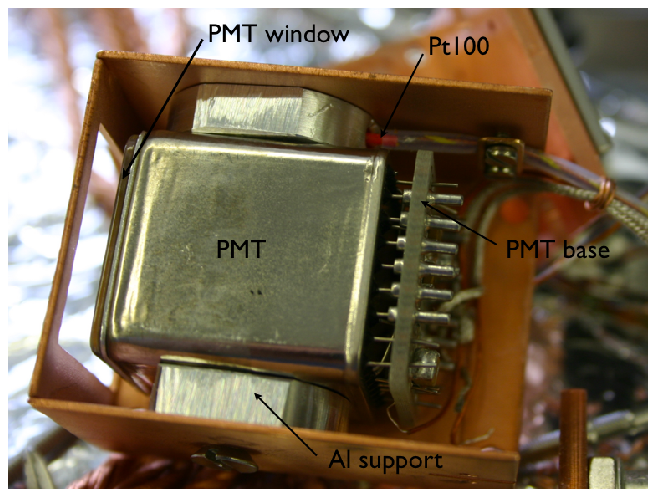}
\includegraphics[width=9cm]{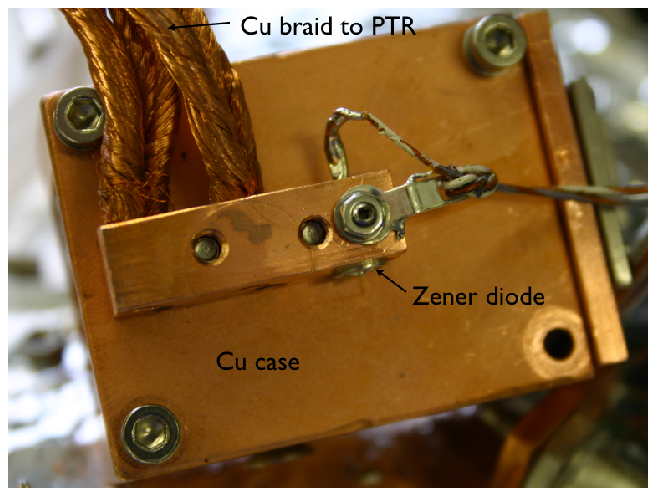}
\caption{The Pt100 mounted on the aluminum support of the PMT (top) and the Zener diode mounted on the copper holding structure (bottom). The copper braid for the cooling of the PMT is also shown.}
\label{PMT_cooling}
\end{center}
\end{figure}

To cool down the PMT from room temperature to a value close to that of LXe ($\sim$165 K), an Iwatani PDC08
Pulse Tube Refrigerator (PTR) is coupled to the vacuum chamber. Two copper braids are connected to the PTR and
attached to each side of a copper box, which is in direct contact with
the case of the PMT. One of two Pt100 thermometers is mounted on an aluminum piece that supports the PMT as can be seen in top of Figure~\ref{PMT_cooling}. Another Pt100 is attached to the PMT wall, allowing a direct measurement of the PMT temperature (not shown in the Figure). 
The case of the PMT is connected to a positive voltage while the PMT base is at the negative voltage. For electrical insulation of the PMT due to the polarity used in the PMT base, the PMT was wrapped around with PTFE tape that also helped hold the Pt100 in place. The signal of the Pt100 is fed into a
customized LabView program and used for a PID control system that
regulates the temperature, via resistive heating of 6.2 V Zener diodes. 
The pressure inside the chamber is kept below 10$^{-6}$ mbar for the duration of all measurements.

An Ortec HV supply provides the high voltage on the
PMT. A modified voltage divider allows a fixed voltage difference to be set between the photocathode and the first dynode in the
recommended linearity range between 50 and
300~V~\citep{Hamamatsu_handbook}. The current from the photodetectors is
measured by a Keithley 6485 picoammeter with a precision of 0.01~pA. Individual measurements of the current, each set to take 1 second, are stored by a dedicated routine provided by
Keithley. For each temperature setting, 100 measurements are taken
consecutively within 2 minutes and are used to determine the mean of the photocurrent and its fluctuations. 

The four PMTs
used in the measurements are Hamamatsu R8520-406 with
bialkali photocathodes and dimensions of 25.7 $\times$ 25.7 $\times$ 28.2 mm$^3$, with an effective photocathode area of 20.5 $\times$ 20.5 mm$^2$. The QE and the Sk$\mathrm{_b}$ at room temperature was provided by
the manufacturer for each of these PMTs. The average QE value is $\sim$31\% at 178 nm Xe
scintillation light~\citep{hamamatsu}. Table~\ref{table_pmt} shows some specifications of this PMT. 

\begin{table}[ht]
\caption{Specifications of R8520-406 PMTs provided by Hamamatsu.}
\begin{center}
\begin{tabular}{c| c}
\hline\hline
Parameter & Description \\
\hline
Spectral Response & 160 to 650 nm\\
Window Material & Synthetic silica \\
Photocathode Material & Bialkali \\
Radioactivity & 15 mBq/PMT Max \\
Operating Ambient Temperature&-110 to +50 deg.C \\
Maximum Supply Voltage & 900 V\\
Pressure-resistance & 5 atm\\
Gain & 1.0$\times$ 10$^6$ \\
Dynode-stage & 10 \\
\hline
\end{tabular}
\end{center}
\label{table_pmt}
\end{table}%

A calibrated silicon photodiode of type AXUV-100G from NIST is used as a
reference detector to measure the light flux at the position where
the PMT is located. The photodiode has an active area of 10 $\times$ 10 mm$^2$. The detector is sensitive to UV light from 5 nm to 254 nm, and its QE has been determined by NIST between
116~nm and 254~nm~\citep{NIST}. This allows us to directly measure the QE of the PMT. 

\section{3. Calibration and Data Taking Procedure}

The monochromator has been calibrated by scanning the wavelength region of a Hg/Cd lamp and comparing the output spectrum with the one provided in the literature~\citep{spetrum_hg}. The relationship between the actual wavelength and the value seen from the monochromator can be described by a simple linear function and was taken into account for selecting the wavelength of interest:

\begin{equation}
\lambda(\lambda_{MP}) = (11.8 \pm 2.2) \AA + (0.9996 \pm 0.0005) \AA \cdot \lambda_{MP} 
\end{equation}
where $\lambda_{MP}$ is the wavelength indicated on the scale of the monochromator and $\lambda(\lambda_{MP})$ is the actual wavelength as a function of $\lambda_{MP}$.

A calibration with the Na-D lines was done with a slit size of 20$~\mu$m each. The resolution of the monochromator is dependent on the size of the entrance and exit slits as well as the wavelength.
This configuration of the monochromator is expected to yield a resolution of 1 $\AA$, which allows the two Na transition lines, from $\lambda(P_{3/2})$ at 5890 $\AA$ and $\lambda( P_{1/2})$ at 5896 $\AA$, to be separated. From the output of the monochromator, two peaks at $\lambda_{MP} = 5788 \AA$ and $\lambda_{MP} = 5794\AA$ can be resolved. 

To measure the QE at room temperature, the number of
photoelectrons measured by the PMT was compared to the number of
photoelectrons measured by the calibrated photodiode. The ratio between
the current read from the PMT and the photodiode is directly
proportional to the ratio of the QE of the PMT to the one of the
photodiode, written as:

\begin{equation}
\mathrm{QE}_{PMT}=\mathrm{QE}_{\mathrm{photodiode}}\times\frac{\mathrm{I}_{PMT}}{\mathrm{I}_{\mathrm{photodiode}}}\\
\label{QE_geo}
\end{equation}
where I$_{PMT}$ and I$_{\mathrm{photodiode}}$ are the currents measured by the picoammeter for the PMT and the photodiode. 

\begin{figure}[htb]
\begin{center}
\includegraphics[width=9cm]{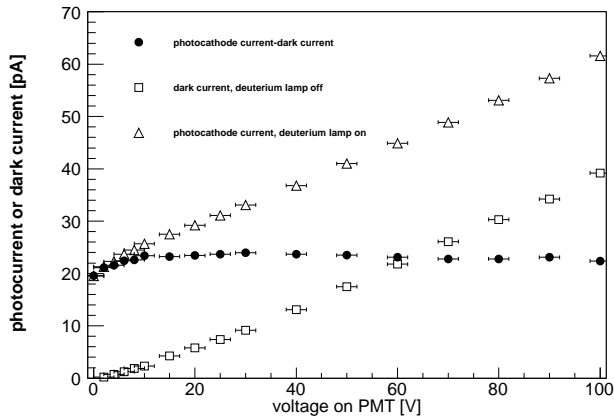}
\caption{Photocathode currents measured as function of the applies voltage: photocathode current deuterium lamp on (triangles), dark current (squares, deuterium lamp off). The circles show the photocurrent induced by the UV photons as the difference of the photocathode current with deuterium lamp on and off.} 
\label{optimal_V}
\end{center}
\end{figure}

The current from the PMT's photocathode is measured when a
voltage is applied between photocathode and the first dynode. The dependence
of the signal size on the voltage is shown in
Figure~\ref{optimal_V}. To correctly estimate the signal of the PMT,
the dark current, due to electron emission from the photocathode not
related to light absorption~\citep{Hamamatsu_handbook}, has to be
subtracted. This is measured at each configuration of the measurement by turning off the lamp and
measuring the current from the PMT. The filled circles in
Figure~\ref{optimal_V} corresponding to the corrected signal show that
the photocurrent is flat for voltages above 20V. A value of 50V, well
within the range of the plateau, was chosen for the final measurements.

\begin{figure}[htb]
\begin{center}
\includegraphics[width=9cm]{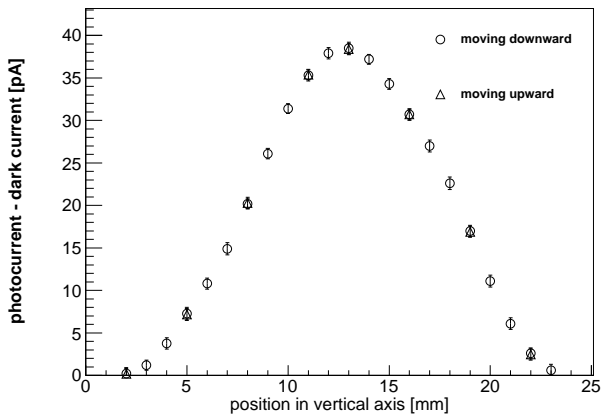}
\caption{Beam profile measured by scanning through
  the vertical axis with the photodiode. Scanning from one direction to
  another (circle) and returning to the original position
  (triangle) measures the same current at each
  position.}
\label{beamProfile}
\end{center}
\end{figure}

To make sure that the beam size at the position of the PMT is smaller
than the area of the photodiode, so that no geometrical corrections for the
flux are needed, the beam profile was scanned in two orthogonal axes
parallel to the surface of the photodiode. The entering slit from the monochromator is equipped with a customized aperture to define the opening of the light beam. Figure~\ref{beamProfile} shows scans along one of the axes. When the photodiode is translated 10 mm away from its center, which is the width of the  photodiode, there is no current reading. Thus, the light flux shining on the photodiode and the PMT is the same, and a geometrical correction is not required to compare these measurements. 

The stability of the emission of the deuterium lamp was monitored for $\sim$12 hours, and the variations of the signal throughout the measurement were found to be
smaller than 1\%. Thus, it is unnecessary to correct the signals from the photodiode and the PMT, which are taken within a few minutes, for such variations.

To reduce the material outgassing, the entire system was baked for
several days while the PMT and the PMT base were kept outside. Because
of some soldered components within the chamber, the temperature was kept below
473 K. Since the chamber had to be opened several times
to change the PMTs, it was either baked again after
the opening at a temperature of $\sim$473 K overnight or was flushed with ultra-high purity N$_{2}$ during the opening of the chamber.

To measure the response of the PMTs at low temperature, it is important
to keep a good vacuum in the chamber in order to prevent both the
absorption of UV as well as the condensation of molecules on the cold PMT quartz 
window. The main contribution to UV absorption are water molecules
with a cross section of $\sigma_{H_2O}$ = 2 $\times$
10$^{-18}$cm$^{2}$ for wavelengths near 178 nm~\citep{Chung_2001}. The contribution from other
contaminants, such as oxygen, is 3 orders of magnitude smaller ($\sigma_{O_2}$ = 4
$\times$10$^{-21}$cm$^{2}$)~\citep{Hudson_1972} at the same wavelength and therefore negligible.

During the cool-down of the chamber, the body of the
PMT is the coldest surface inside the vacuum chamber
and water from the residual gas may condense on its 
surface and cause light absorption. 
At our lowest temperature of $T_{min}=160$~K,
the saturation pressure of water decreases down to $\approx 10^{-6}$~mbar~\citep{nist_sat}.
Since we kept the total pressure inside the vacuum chamber always below 
$10^{-6}$~mbar and the partial pressure of water in the chamber was even lower, 
no condensation of water could have happened at the quartz window of the PMT. 
Of course, these arguments only hold for condensing of a bulk film, 
which would be governed by the adsorbate-adsorbate (water-water) interactions. 
The arguments do not contradict a possible monolayer of water on the quartz 
window of the PMT, which would be governed by adsorbate-substrate 
(water-quartz) interactions. But even a monolayer of water would not have influenced
the QE measurement significantly.
Using the typical surface density of a monolayer of $nd = 10^{15}$~cm$^{-2}$, 
the absorption probability amounts to 
$P_{abs} = nd \cdot \sigma_{H_2O} = 2 \cdot 10^{-3}$ only and would have changed
out QE determination by 0.2\%.
This possible systematic effect is a factor 10 smaller than the other uncertainties.

For a better understanding of the impurities in the system, an Extorr
XT300 Residual Gas Analyzer (RGA) was installed to monitor
the water peak as the system was cooled down, and a cryopump
was added to effectively trap gas that could affect the measurements at low temperature. The residual gas content was monitored by computer with a
continuous scan of the whole mass range. The amount of water decreased below $\sim$270K, which
can be explained by water freezing around that temperature. Although the dark current drops below 270K, the signal, i.e., the
difference between photocurrent and dark current, does not change
within the measured fluctuations. If water were sticking to
the PMT window, a decrease in the response to the light would have
been observed. This demonstrates that the impact of water on our results is negligible. 

\section{4. Results and Discussions}
\subsection{4.1. QE measurements at room temperature}

As a first step, the QE at room temperature is measured for the four PMTs and the results are compared with the values provided by
Hamamatsu. Using as a reference the NIST photodiode, as explained in
the previous section, the photocurrent from each PMT is measured and
transformed into a QE according to
Eq.~\ref{QE_geo}. Table~\ref{table_qe} shows the values provided by
Hamamatsu and the values measured with the present setup. For both the photomultipliers and the reference photodiode, the standard deviation of the fluctuations in each dataset was taken as the uncertainty on the measurements. 

\begin{table}[ht]
\caption{QE values measured at room temperature and comparison with
  the values provided by Hamamatsu for 4 different PMTs, designated by their serial numbers}
\begin{center}
\begin{tabular}{c c c}
\hline\hline
PMT S/N & Hamamatsu QE [\%] & Measured QE [\%] \\
\hline
LV1002 & 30.5  & 31.1 $\pm$ 2.7\\
LV1009 & 30.0  & 31.3 $\pm$ 5.0\\
LV1013 & 30.9  & 31.6 $\pm$ 1.5\\
LV1014 & 30.3 & 31.3 $\pm$ 3.6 \\
\hline
\end{tabular}
\end{center}
\label{table_qe}
\end{table}%

\subsection{4.2. QE measurements at low temperature}
To measure the QE of the PMT as a function of temperature, the PMT
casing is cooled down with the PTR. The temperature is kept at its set point by heating with the Zener diode. Since the geometry inside the chamber remains unchanged, the amount of light incident on the PMT
window stays the same and the QE at low temperature can be inferred from
the room temperature measurement according to:

\begin{equation}
QE(T)=\frac{I(T_0)}{I(T)}QE(T_0)
\end{equation}

\begin{figure}[htb]
\begin{center}
\includegraphics[width=9.5cm]{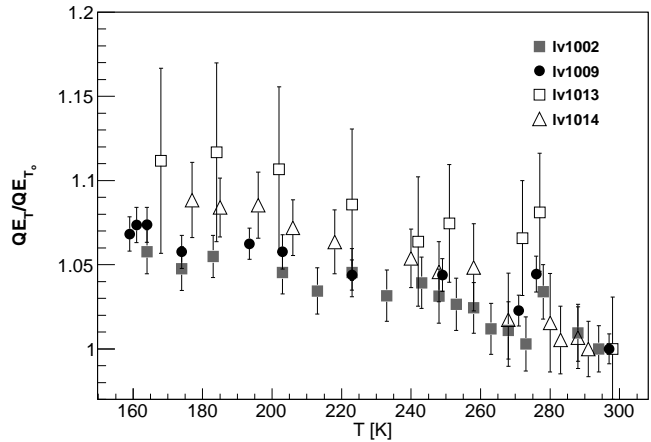}
\caption{The dependence of QE on temperature, shown for 4 high QE
  PMTs. Each color corresponds to a different PMT, identified by its
  serial number in the legend.}
\label{relative}
\end{center}
\end{figure}

Thus, no further comparison to the photodiode is necessary. Current measurements were performed from room temperature down to 160 K, the
lowest operation temperature of the PMTs, as indicated by the
manufacturer. Figure~\ref{relative} shows the ratio between the QE at
low and room temperature for the four PMTs studied, as a function of
temperature. Note that the room temperatures for each PMT are slightly
different, as shown by the rightmost points in each curve, thus
shifting the reference point for each. 

For all the PMTs, the sensitivity increases up to $\sim$10\% at the lowest
temperature that was measured. No hysteresis effect was
observed while warming back up to room temperature in any of these measurements. The uncertainties reflect the standard deviation values of 100 current measurements at each
temperature. The larger error bars for PMT LV1013 are due to electronic noise from running the turbo pump during that measurement.
 
\section{5. Conclusion}
\label{}
We have designed a system to measure the QE of
R8520-406 PMTs, which have special bialkali photocathodes capable of
providing a good performance at temperatures down to
160~K. Four PMTs were measured with this setup. Their absolute QEs
at room temperature were measured by comparing their response to that of a calibrated photodiode, and the values obtained match those quoted by the manufacturer within the uncertainties. 

The QE of the PMTs was also measured for temperatures down to
160 K, the temperature of LXe at which these PMTs are usually operated. A relative increase at low
temperatures of up to $\sim$10\% is measured for all four PMTs. This constitutes an
important measurement for detectors based on LXe scintillation since it
will help to more accurately determine the number of photons observed from the number of photoelectrons detected. The setup can be adopted to enable testing of other types of PMTs being considered for next generation experiments based on liquid noble elements.

\vspace{1cm}
\section{Acknowledgements}

This work was carried out with support from the National Science Foundation and from a start-up grant of the University of M\"unster. We gratefully acknowledge the help from NIST and from Robert E. Vest, in particular for valuable advice and the opportunity to use one of their photodiodes. 
\vspace{-.8cm}


\end{document}